\documentclass[aps,prl,amsmath,amssymb,reprint,superscriptaddress]{revtex4-1}

\usepackage{graphicx}
\usepackage{xcolor}

\newcommand{\whbar}{\tilde \hbar}
\newcommand{\gave}{\langle g \rangle}

\begin{document}

\title{Enhancement of many-body quantum interference in chaotic bosonic systems: the role of symmetry and dynamics}

\newcommand{\RegensburgUniversity}{Institut f\"ur Theoretische Physik, Universit\"at Regensburg, D-93040 Regensburg, Germany}
\newcommand{\LiegeUniversity}{CESAM research unit, University of Liege, 4000 Li\`ege, Belgium}
\newcommand{\Steve}{Department of physics and astronomy, Washington State University, Pullman, WA USA}
\newcommand{\LPTMS}{LPTMS, CNRS, Univ. Paris-Sud, Universit\'e Paris-Saclay, 91405 Orsay, France}

\author{Peter Schlagheck}
\email{Peter.Schlagheck@liege.be}
\affiliation{\LiegeUniversity}
\author{Denis Ullmo}
\affiliation{\LPTMS}
\author{Juan Diego Urbina}
\affiliation{\RegensburgUniversity}
\author{Klaus Richter}
\affiliation{\RegensburgUniversity}
\author{Steven Tomsovic}
\affiliation{\RegensburgUniversity}
\affiliation{\Steve}

\begin{abstract}

Although highly successful, the truncated Wigner approximation (TWA) leaves out many-body quantum interference between mean-field Gross-Pitaevskii solutions as well as other quantum effects, and is therefore essentially classical.  Turned around, if a system's quantum properties deviate from TWA, they must be exhibiting some quantum phenomenon, such as localization, diffraction, or tunneling.  Here, we examine a particular interference effect arising from discrete symmetries, which can significantly enhance quantum observables with respect to the TWA prediction, and derive an augmented TWA in order to incorporate them. Using the Bose-Hubbard model for illustration, we further show strong evidence for the presence of dynamical localization due to remaining differences between the TWA predictions and quantum results.

\end{abstract}



\maketitle

For an ultracold bosonic system sufficiently isolated from its environment, a host of fascinating many-body quantum phenomena can be explored.  Effects such as various aspects of quantum tunneling~\cite{Hensinger01,Folling07}, quantum revivals of coherent states that have initially dispersed~\cite{Greiner02b}, and the more subtle coherent backscattering in Fock space~\cite{Engl14b,Engl18}, are all significant examples.

In the context of single particle or few-body systems, some phenomena have their most striking manifestation in a system's time evolution, such as the aforementioned revivals~\cite{Yeazell90,Robinett04}, but others, such as localization in its various forms, also have unique signatures in the time domain~\cite{Fishman82,Shepelyansky83,Heller84,Ketzmerick92,Bohigas93}.  The onsets of these signatures typically begin beyond an Ehrenfest time scale, $\tau_E$,~\cite{Ehrenfest27, Berman78} in which even the most localized initial states must disperse.  For bounded, strongly chaotic dynamical systems, this time scale is logarithmically-short-in-$\hbar$.
 
The high density limit of ultracold bosonic many-body systems has a deep formal similarity to the semiclassical limit of few-body systems, with the inverse filling factor playing the mathematical role of $\hbar$, and the classical mean-field solutions (of the Gross-Pitaevskii (GP) equation or its discrete version) the role of classical trajectories~\cite{Engl14b}. A complete semiclassical theory includes interference, diffraction, and tunneling, and thus, generally speaking, quantum phenomena identified in few-body systems are expected to have their analogs in many-body systems. This, in particular, implies a breakdown beyond $\tau_E$ of the ``naive'' classical mean-field approximation.  It is far from evident though how such post-Ehrenfest processes can be employed in practice for exploring genuinely quantum effects associated with bosonic many-body systems, such as interferences in Fock space.  Indeed, although these interferences are supposed to manifest themselves sensitively in the form of rapid oscillations within the quantum many-body wavefunction's time evolution, the observable impacts of these rapid oscillations typically are effectively washed out if the detailed information contained in the many-body wavefunctions is projected onto a generic single-particle operator's expectation value.

This actuality, discussed below in greater detail, lies at the heart of the success of the essentially classical phase-space method known as the truncated Wigner approximation (TWA)~\cite{Steel98,Sinatra02,Polkovnikov10}.  In practice, TWA amounts to performing a Monte-Carlo sampling of the time evolution of a quantum many-body state in terms of GP trajectories, i.e., classical fields that evolve according to a GP equation, whose initial values are randomly chosen such that they correctly sample the phase space distribution of the initial quantum state under consideration~\cite{Steel98,Sinatra02}.  The possible occurrence of quantum interference between those GP trajectories is completely neglected within the TWA (as are diffractive or classically prohibited trajectories), more or less as though the systems were weakly, but just sufficiently, coupled to a decohering environment.

As long as some effective time average is performed when comparing with experimental data, there is a general expectation that the above reasoning remains valid also for the expectation values of more sophisticated many-body observables, such as the detection probability of a given Fock state with respect to a single-particle basis.  Conversely, a significant deviation of a system's time-averaged behavior with respect to the TWA prediction is a sensitive indicator of some special many-body quantum phenomenon.  

Ahead, coherent state survival probabilities, i.e.~the absolute square of an evolved coherent state's overlap with it's initial self, are considered as an especially interesting class of measures.  The coherent states have minimum uncertainty and are initially the most classical possible~\cite{Glauber63}.  Furthermore, the survival probability, with some time averaging that eliminates the ``generic'' rapid interference oscillations contains a great deal of information about the various other surviving and more robust forms of quantum effects.  Discrete symmetries, which have a long history, e.g.~from Bragg scattering~\cite{Martin88}, molecular spectroscopy~\cite{Harris89}, electronic transport~\cite{Baranger91}, to topological materials~\cite{Ando13}, lead to such surviving constructive interferences.  Here we show how to incorporate those effects into the TWA, and how such symmetry-related interferences emerge after $\tau_E$. We also argue that the further deviations after time averaging between the TWA and certain Bose-Hubbard model cases shown ahead are signatures of dynamical localization due to quantum interference in a many-body context. 

Clearly, for the survival probability an excess of constructive interference for any reason would lead to the TWA underestimating the quantum results, and conversely, an excess of destructive interference would lead to the TWA overestimating them.  It is not immediately obvious how to incorporate dynamical effects into the TWA such as scarring~\cite{Heller84}, dynamical localization in systems with classical transport barriers~\cite{Bohigas93} or diffusive dynamics~\cite{Fishman82}, or tunneling, although such effects would create telltale signals in the behavioral differences between the TWA and quantum results~\cite{Ullmo20}.  However, it is possible to derive an enhanced version of the TWA accounting for {{\it discrete symmetry quantum interference  effects}.  

\begin{figure}[t]
\begin{center}
\includegraphics[width=\columnwidth]{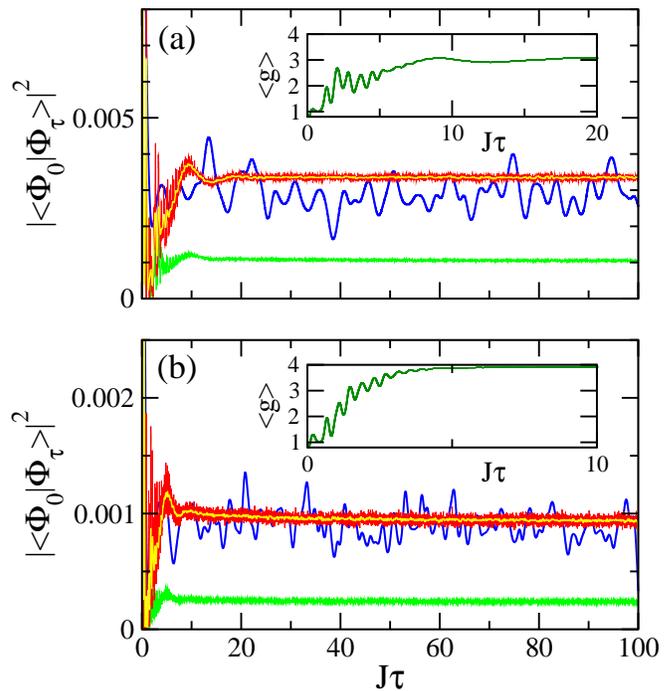}
\end{center}
\caption{(Color online) (a) Survival probability of an initial condensate coherent state, $|\Phi_0\rangle$, versus scaled time $J\tau$.  The time averaged~\cite{rem_av} quantum calculation, dark (blue) line, is done for an initial coherent state centered at $(\sqrt{9},1,\sqrt{9},1)$ for an $L=4$-site Bose-Hubbard ring (hopping and on-site interaction energies, $J$ and $U$, respectively, and $N=20$ the mean total number of particles) with  $NU/LJ = 0.5$.  Conventional TWA simulations, noisy (green) line, significantly underestimate the quantum result.  The symmetry enhanced TWA, noisy (red) line with its time average~\cite{rem_av} (yellow line) on top,  is much closer. The insert shows the ratio of the two TWA calculations, $\gave$.  (b) Similar to (a) except for the coherent state centered at $(\sqrt{18},\sqrt{2},\sqrt{18},\sqrt{2})$ for which  $\tilde \hbar$ is half as large.  The agreement improves and $\gave$ is seen to approach the maximum possible, $L=4$, unlike in (a) where it approaches $\approx 3.2$ (see insets).  
\label{fig:s4}}
\end{figure}

Indeed, as illustrated in Fig.~\ref{fig:s4}, if both the Hamiltonian and the initial state of the survival probability are symmetric individually with respect to some group of symmetry operations (e.g., reflection, permutation, discrete rotation), then the usual TWA calculation underestimates the mean value of such an observable beyond $\tau_E$ by a time-dependent mean symmetry factor $ \gave$.  The unaccounted for enhancement is due to systematic constructive many-body interferences that arise between the contributions of symmetry-related GP trajectories.  This enhancement is similar to phenomena found in a single particle physics context~\cite{Baranger91,Oconnor91}.  However, with increasing numbers of degrees of freedom comes the possibility of much larger discrete symmetry groups where this factor can become arbitrarily large. 

The TWA can be obtained from a semiclassical van Vleck-Gutzwiller description of the time evolution of ultracold bosonic system by applying the diagonal approximation~\cite{Berry85,Sun98,Dittrich06,Dujardin15b,Tomsovic18}.  To show this, we assume that the time evolution can be well represented within a finite discrete one-body basis containing $L$ single-particle wavefunctions.  In the case of a Bose-Einstein condensate that is confined within an optical lattice, those single-particle wavefunctions would most naturally be given by the Wannier orbitals associated with the individual lattice sites.  The quantum system is generally described by a Bose-Hubbard Hamiltonian,
\begin{equation}
\hat{H} = \sum_{l,l'=1}^L H_{l,l'} \hat{b}_l^\dagger\hat{b}_{l'} 
+  \frac{1}{2} \sum_{l=1}^L  U_l \hat{b}_l^\dagger\hat{b}_l^\dagger\hat{b}_l\hat{b}_l
\label{eq:bh}
\end{equation}
with $H_{l,l'}=H_{l',l}^*$ the one-body Hamiltonian matrix elements and $U_l$ the on-site interaction energies.

We introduce the quadrature operators $(\hat q_j, \hat p_j)$ defined as 
\begin{equation} 
\label{eq:quadratures}
\hat q_j + i \hat p_j = \sqrt{2 \whbar} \hat b_j  \quad , \quad \hat q_j - i \hat p_j = \sqrt{2\whbar} \hat b_j^\dagger \; ,
\end{equation}
where $\whbar$ equals the mean filling factor inverse, i.e.~ratio of the site number to mean total particle number, $L/N$. Following Ref.~\cite{Dujardin15b}, we represent the semiclassical propagator in the $| \mathbf{q}\rangle$ basis associated with $\mathbf{\hat q} \equiv (\hat q_1,\ldots,\hat q_L)$.  For a large total particle number, the time evolution operator matrix elements can be expressed within the quadrature basis as a sum over mean-field trajectories \cite{Littlejohn92},
\begin{equation}
\langle \mathbf{q}_f | e^{-i\tau\hat{H}/\whbar} | \mathbf{q}_i\rangle = \sum_\gamma A_\gamma(\mathbf{q}_f,\mathbf{q}_i) e^{i R_\gamma(\mathbf{q}_f,\mathbf{q}_i)/\whbar {- i \kappa_\gamma \pi/2}} \,
\label{eq:gp}
\end{equation}
which go from $\mathbf{q}_i$ to $\mathbf{q}_f$ in the scaled time $\tau=\tilde \hbar t/\hbar$. For the Hamiltonian of Eq.~\eqref{eq:bh} such a trajectory would correspond to a solution of the GP equation
\begin{equation}
i \whbar \frac{\partial}{\partial \tau} \psi_l(\tau) = \sum_{l'=1}^L H_{l,l'}\psi_{l'}(\tau) + U_l(|\psi_l(\tau)|^2 - 1)\psi_l(\tau)
\end{equation}
satisfying the boundary conditions $\mathrm{Re}[\psi^l(0)] = q^l_i/\sqrt{2 \whbar}$ and $\mathrm{Re}[\psi^l(\tau)] = q^l_f/\sqrt{2 \whbar}$. $R_\gamma(\mathbf{q}_f,\mathbf{q}_i)$ represents its associated principal function or action integral. The integer Morse index $\kappa_\gamma$ \cite{Littlejohn92} counts the number of conjugate points along the trajectory~\footnote{See supplemental material for further details.}.

Any generic single- or many-particle observable in this bosonic system can be expressed in terms of the quadrature operators as $\hat{f} \equiv f(\hat{\mathbf{q}}, \hat{\mathbf{p}})$ yielding, e.g., $\hat{b}_l^\dagger\hat{b}_l = (\hat{q}_l^2 + \hat{p}_l^2 - \whbar)/2\whbar$ for the occupancy on site $l$.  Given an initial many-body state $|\Phi_0 \rangle$, the time evolution $\langle\hat{f}\rangle(\tau) \equiv {\rm Tr} [ \hat f \hat \rho_{\tau}] $ of the mean value of such an observable, with $ \hat \rho_\tau \equiv \vert\Phi_\tau\rangle \langle\Phi_\tau \vert$ and $\vert\Phi_\tau\rangle \equiv e^{-i\tau\hat{H}/\whbar} \vert \Phi_0 \rangle$, is then expressed as
\begin{eqnarray}
&&\langle\hat{f}\rangle(\tau)  =  \int d\mathbf{q}_i'd\mathbf{q}_i'' d\mathbf{q}_f'd\mathbf{q}_f'' \Phi_0^*(\mathbf{q}_i') \Phi_0(\mathbf{q}_i'') \langle \mathbf{q}_f' \vert \hat f \vert \mathbf{q}_f''\rangle \nonumber \\&\times & \sum_{\gamma' : \mathbf{q}_i' \to \mathbf{q}_f'} \sum_{\gamma'' : \mathbf{q}_i'' \to \mathbf{q}_f''} A_{\gamma'}^*(\mathbf{q}_f',\mathbf{q}_i') A_{\gamma''}(\mathbf{q}_f'',\mathbf{q}_i'') \label{eq:sc} \\&\times& \exp \left[ \frac{i}{\whbar} (R_{\gamma''}(\mathbf{q}_f'',\mathbf{q}_i'') - R_{\gamma'}(\mathbf{q}_f',\mathbf{q}_i')) - i(\kappa_{\gamma''} - \kappa_{\gamma'}) \frac{\pi}{2} \right] \; ,  \nonumber
\end{eqnarray}
where $\Phi_0(\mathbf{q})$ represents the initial quantum many-body wavefunction in the $q$-quadrature basis.

For an averaging process that suppresses oscillating terms, the main contributions to this integral are expected to be given by the diagonal approximation.  This implies two crucial assumptions:  (i) that in the double sum over trajectories, all terms for which the two action integrals $R_{\gamma'}$ and $R_{\gamma''}$ correspond to different orbits cancel out; and (ii) only {\em short chords}~\cite{Ozorio98}, i.e.\ points such that $\mathbf{q}_i' \simeq \mathbf{q}_i''$ and $\mathbf{q}'_f \simeq \mathbf{q}_f''$ are going to contribute significantly to the integrals.  If so, one may transform the double sum in Eq.~\eqref{eq:sc} into a single sum over mean field trajectories and expand in the small parameter associated with the chord.  This leads to
\begin{equation} 
\label{eq:TWA}
\langle\hat{f}\rangle_{\rm diag}(\tau) = \int \frac{d\mathbf{Q}_i d\mathbf{P}_i}{(2\pi \whbar)^L}   [\rho_0]_W ( \mathbf{Q}_i, \mathbf{P}_i) \, [f]_W ( \mathbf{Q}_f, \mathbf{P}_f) \; , 
\end{equation}
where $(\mathbf{Q}_f, \mathbf{P}_f)$ has to be understood as the final point in phase space at time $\tau$ of the trajectory initiated at $(\mathbf{Q}_i, \mathbf{P}_i$), and  with the Wigner transform of an operator $\hat O$ defined as 
\begin{equation} 
[O]_W (\mathbf{Q}, \mathbf{P}) \equiv \int d\mathbf{\delta q} e^{(i/\whbar)\mathbf{P}\cdot\mathbf{\delta q} } \langle \mathbf{Q} + \frac{\mathbf{\delta q}}{2} \vert \hat{O} \vert \mathbf{Q} - \frac{\mathbf{\delta q}}{2} \rangle \; .
\end{equation} 
Provided $f$ is a rather smooth and well-behaved function of $\mathbf{q}$ and $\mathbf{p}$, $[f]_W ( \mathbf{Q}_f, \mathbf{P}_f)$ can, to within small $\whbar$ corrections, be approximated by $f(\mathbf{Q}_f, \mathbf{P}_f)$, and the integral in Eq.~\eqref{eq:TWA} can be evaluated by a Monte-Carlo method. This becomes the TWA.

The above reasoning would be valid if either the expectation value itself, or some further temporal or configuration averages, removes the contributions of Eq.~\eqref{eq:sc}'s off-diagonal terms.  However, if there exists a discrete symmetry, more care must be exercised.  Consider an initial condensate wavefunction that is symmetric with respect to a given parity exhibited by the Hamiltonian and a many-body observable $\hat f$. Within the phase space there is a subspace of points that are their own parity partners. Denote it the symmetry subspace.  At long times nearly every trajectory $\gamma$ that significantly contributes to the semiclassical expression, Eq.~\eqref{eq:sc}, for the mean value of this observable leaves the neighborhood of the symmetry subspace and has a symmetry-related partner trajectory $\tilde{\gamma}$, which is obtained by applying the parity operator onto $\gamma$ and which exhibits the same action integral as $\gamma$.  In particular, partner trajectories satisfy $R_{\tilde{\gamma}}(\tilde{\mathbf{q}}_{{f}}, \tilde{\mathbf{q}}_{{i}}) = R_\gamma(\mathbf{q}_{{f}},\mathbf{q}_{{i}})$ and thus also $A_{\tilde{\gamma}}( \tilde{\mathbf{q}}_{{f}},\tilde{\mathbf{q}}_{{i}}) = A_\gamma(\mathbf{q}_{{f}},\mathbf{q}_{{i}})$. The contributions of those two trajectories therefore interfere constructively within Eq.~\eqref{eq:sc}.  As a consequence, Eq.~\eqref{eq:TWA}, which entirely neglects those interferences, underestimates the true expectation value of $\hat f$ by a factor two.

To account for this effect in a quantitatively correct manner within TWA, each trajectory must be associated with its particular symmetry factor, $g_\gamma$, that correctly counts the number of other symmetry-related trajectories with which constructive interference will arise.  A naive way to approach that problem would be to perform the Monte-Carlo calculation implied by Eq.~\eqref{eq:TWA} and multiply each contribution by its number of existing distinct symmetry related orbits.  However, since the number of orbits that are symmetric under an element of the symmetry group is of measure zero, this would lead to multiplying the TWA result by a global factor $g_{\rm max}$ (the number of symmetry group elements), which entirely misses the transient shift from unity to maximum at relatively short dynamical times, and misses the fact that due to coherent state spread, the long time $\gave$ can saturate below $g_{\rm max}$; see Fig.~\ref{fig:s4}a where $g_{\rm max}=4$.  

A less intuitive and more precise picture for why this approach cannot succeed is that the correct enhancement factor $g_\gamma$ is not the symmetry of the orbit itself, but that of its family of orbits defined by the neighborhood of a ``saddle trajectory''~\cite{Tomsovic18b}.  This aspect appears when some of the integrals of Eq.~\eqref{eq:sc} are performed within the stationary phase approximation (hence, ``saddle trajectory''), and the contributions of orbit families (saddle trajectory neighborhoods) are naturally grouped together.  This grouping ensures that orbits within a family are similar for their entire evolution, and in particular, are well represented approximately through the use of the saddle trajectory's stability matrix.  If an orbit and a symmetric version are within the same family, there is no enhancement.  If they are associated with well separated saddles, which cannot occur in the dynamics for times shorter than $\tau_E$, then there is an enhancement.  Note that the stationary phase approximation involves locating complex saddle trajectories~\cite{Huber87,Huber88,Pal16,Baranger01,Tomsovic18,Tomsovic18b} in order to evaluate the mean value of observables, and lacks the simplicity of the ``initial value'' formulation of the TWA, Eq.~\eqref{eq:TWA}.  

Nevertheless, the conventional implementation of the TWA can be amended to incorporate the impact of discrete symmetries in an automated manner.  First, note that by the saddle logic above, the ``symmetric'' trajectories with respect to a given symmetry operation that do not lead to enhancement must remain during their entire evolution in the neighborhood of the relevant symmetry subspace. Those non-symmetric ones leading to an enhancement factor do not.  Hence, in order to discriminate between those possibilities, it is sufficient to introduce an $\whbar$ distance scale $d$ in phase space through which the notion of ``closeness'' to a symmetry subspace can be properly defined.  A trajectory that is computed within the framework of Eq.~\eqref{eq:TWA} is symmetric with respect to a given symmetry subspace if its distance to that subspace remains below $d$ for its entire evolution time.  Otherwise, it is non-symmetric with respect to that subspace, and in that case its contribution is multiplied by the associated enhancement factor.  Up to $\tau_E$, trajectories cannot leave a symmetric subspace and return. The mean value of $f$ is then provided by symmetric trajectories, i.e., pairs, both of whose members are included within the short chord approximation: their $\gamma$ does not give rise to a distinct $\tilde{\gamma}$ ($\gamma =\tilde{\gamma}$).  Thus, the the dynamics need time to explore phase space sufficiently to manifest the symmetry's existence, and this leads to time scales associated with increasing multiplicities, depending on the symmetry group.

The ``symmetry-enhanced'' curves in Fig.~\ref{fig:s4} have been constructed following this approach.  It shows a homogeneous $1D$ Bose-Hubbard chain with $L=4$ sites and periodic boundary conditions.  This system is modeled by Eq.~\eqref{eq:bh} with the specific choices $U_l = U > 0$ for all $l$ and $H_{l l'} = - J < 0$ if $l' = l \pm 1$ (mod $L$) and zero otherwise.  Furthermore, consider an initial coherent state for a perfect BEC centered about the field amplitudes $\psi_l^{(0)} = (q_l^{(0)} + ip_l^{(0)})/\sqrt{2\whbar}$~\footnote{The breaking of $U(1)$ symmetry that is implied by this ansatz does not play a significant role for most commonly considered observables, provided the BEC contains a reasonably large number $N\gg1$ of particles~\cite{Lieb07}}, which is equivalent to a wave packet up to a global phase~\cite{Tomsovic18}
\begin{equation}
\Phi_0(\mathbf{q}) = \frac{1}{(\pi\whbar)^{\frac{L}{4}}} \exp\left[{-\frac{(\mathbf{q}-\mathbf{q}^{(0)})^2}{2\whbar} + \frac{i}{\whbar} \mathbf{p}^{(0)} \cdot(\mathbf{q}-\mathbf{q}^{(0)})}\right] . 
\label{eq:bec} 
\end{equation}

 More specifically, consider the coherent state density waves: (a)  $(\sqrt{9},1,\sqrt{9},1)$ and (b) $(\sqrt{18},\sqrt{2},\sqrt{9},\sqrt{2})$ as initial states, and  the many-body observable of interest given by $\hat{f} = \vert \Phi_0\rangle\langle\Phi_0\vert = \hat \rho_0$, whose expectation value is the survival probability, $|\langle \Phi_0| \Phi_\tau\rangle |^2$, after a given evolution time $\tau$, where we perform a time average \cite{rem_av} in order to filter out rapid oscillations arising due to the generic quantum interferences. The symmetry group in this example is larger than the one associated with simple parity.  As a consequence, the enhancement factor, which depends on  the group, is larger than two. Indeed, for a Bose-Hubbard ring with equal on-site interaction energies (all $U_l=U$) and only equal nearest neighbor one-body terms, the full discrete group includes cyclic permutations and a clockwise-counterclockwise equivalence.  The maximum possible enhancement factor for such a system, before accounting for the symmetry of the initial state and observable, is twice the number of sites, $2L$.  However, the density wave partially breaks the full system dynamical symmetry as only even cyclic permutations leave it invariant.  Thus, the maximum symmetry enhancement factor is reduced by a factor two to $L$, which equals $4$ in our figures.  

The curve thus obtained accounting for $g_\gamma$ (noisy red line) is seen in Fig.~\ref{fig:s4} to perfectly follow the smoothed exact quantum result.  Evidently, this approach can straightforwardly be generalized to account for the presence of larger symmetry groups in the Hamiltonian and the initial state.  In calculations not shown with $L=6$, the enhanced TWA saturated at a factor $\gave= 6$, and the agreement with the smoothed exact quantum result is excellent.  Note that this factor could be much larger.   For the fully connected Bose-Hubbard model studied in Ref.~\cite{Pizzi19}, the maximum enhancement factor for a density wave survival probability would be $(L/2)!^2$.

\begin{figure}[t]
\begin{center}
\includegraphics[width=\columnwidth]{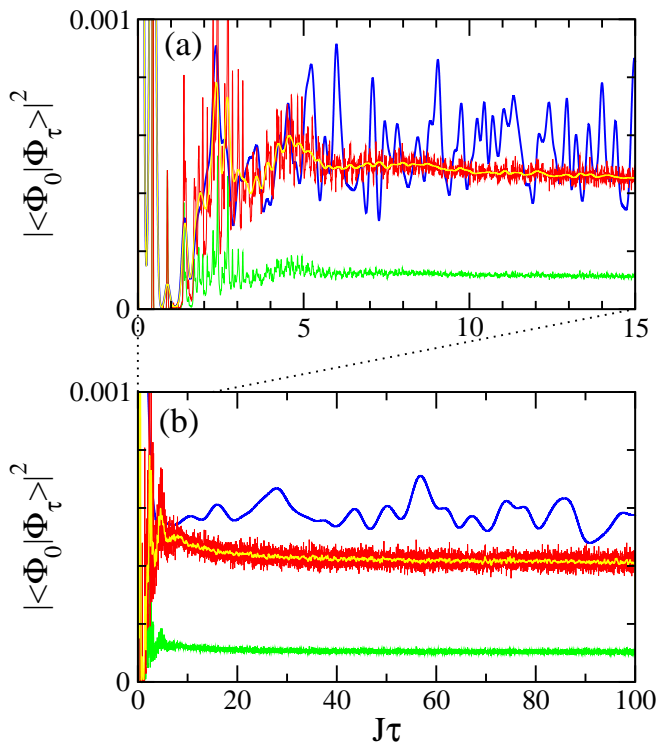}
\end{center}
\caption{(Color online) Same as Fig.~\ref{fig:s4} for the initial condensate coherent state centered at $(\sqrt{19},1,\sqrt{19},1)$ with $NU/LJ = 0.5$.  The slow decay of classical survival probability after $J\tau\approx 10$ is not followed by the quantum time averaged survival probability.  This behavior strongly suggests a dynamical localization process.
\label{fig:s6}}
\end{figure}

In a second example shown in Fig.~\ref{fig:s6}, a different initial coherent state is chosen, other parameters the same as before, for which just beyond $\tau_E$ there is rough agreement between the symmetry enhanced TWA and the smoothed quantum survival probability.  However, with increasing time there is a decrease in the TWA indicating that new regions of phase space are opening up with time, and yet the quantum average does not follow this decrease.  This demonstrates that the system is undergoing additional quantum interference effects, which in this case strongly suggests a dynamical localization process.  One for which there are time scales beyond the Ehrenfest time in which the classical trajectories are continuing to access new regions of phase space that the quantum system cannot.

In summary, the very successful TWA approximation can be extended to account quantitatively for basic constructive interference effects enforced by the existence of discrete symmetries, which can involve large enhancement factors.  In the time domain, discrete symmetries reveal themselves over time and depending on the dynamics incorporate multiple time scales.  For example, the $L=8$ ring has subspaces with factors, $1,2,4,8$.  For some initial condensate coherent states and values of $NU/LJ$, the Lyapunov exponent in the direction of the double degeneracy symmetry subspace is much greater than that for $4$ or $8$.  The factor $2$ then enters the dynamics well before the higher degeneracy factors.

In addition, deviations between the TWA and quantum behaviors of the survival probability indicate the presence of some genuine quantum effect such as localization, diffraction, or tunneling.  In the case shown here, the initial agreement followed by a divergence appears to be due to a quantum interference effect leading to dynamical localization.  This is distinctly different from a classical localization in which the GP solutions are trapped or fail to explore phase space regions common in mixed phase space systems, for example, the macroscopic self-trapping discussed in Ref.~\cite{Pizzi19}. More work is needed to separate signals of the various forms of localization from each other or from tunneling and diffraction effects, and is left for the future.

\acknowledgments

We thank Mathias Steinhuber and R\'emy Dubertrand for scientific discussions and Deutsche Forschungsgemeinschaft (project Ri681/14-1), Vielberth Foundation and Bayerisch-Franz\"osisches Hochschulzentrum / Centre de Coop\'eration Universitaire Franco-Bavarois (project FK56\_15) for financial support.  One of the authors (ST) thanks the Laboratoire de Physique Th\'eorique et Mod\`eles Statistiques for its hospitality and support.

\section*{Supplementary material}

\newcommand{\eqrefEqTWA}{(6)}
\newcommand{\eqrefEqsc}{(5)}

In this supplementary material we provide some of the details leading to Eq.~\eqrefEqTWA.  First, note that the prefactor can be written as
\begin{equation} 
\label{eq:Agamma}
A_\gamma(\mathbf{q}_f,\mathbf{q}_i) = \frac{e^{i \pi \alpha/4}}{(2\pi \whbar)^{L/2}} \left|\det\left(-\frac{\partial^2 R_\gamma}{\partial q^l_f \partial q^{l'}_i} (\mathbf{q}_f,\mathbf{q}_i)\right)\right|^{1/2}
\end{equation}
with $\alpha$ the index of inertia of $-\partial \mathbf{q}_i/\partial \mathbf{p}_i$ \cite{Littlejohn92}.  Next, starting from Eq.~\eqrefEqsc, assume that the contributions of non-identical pairs of orbits cancel out because of the time average, and that only short chords, i.e.\ points such that $\mathbf{q}_i' \simeq \mathbf{q}_i''$ and $\mathbf{q}'_f \simeq \mathbf{q}_f''$ contribute significantly.  In such a case, the variables $\mathbf{Q}_{f,i} \equiv (\mathbf{q}_{f,i}'' + \mathbf{q}_{f,i}')/2$ and  $\mathbf{\delta q}_{f,i}  \equiv (\mathbf{q}_{f,i}'' - \mathbf{q}_{f,i}')$ can be introduced.  The key is to expand in the ``small'' variables $\mathbf{\delta q}_{f,i}$, which means more specifically keeping only the zero'th order term for the smooth function $A_\gamma$ but expanding the action $R_\gamma$ to first order.  With $\gamma'=\gamma''=\gamma$, and using the property $\mathbf{p}^{(\gamma)}_f = \partial R_\gamma/\partial \mathbf{q}_f$; $\mathbf{p}^{(\gamma)}_i = - \partial R_\gamma/\partial \mathbf{q}_i$ (with $\mathbf{p}^{(\gamma)}_i$ and $\mathbf{p}^{(\gamma)}_f$ the initial and final ``momenta'' of the trajectory $\gamma$),  gives
\begin{eqnarray}
&&\langle\hat{f}\rangle(t)_{\rm diag}  =  \int d\mathbf{Q}_i d\mathbf{Q}_f d\mathbf{\delta q}_f d\mathbf{\delta q}_i \, \langle \mathbf{q}_i''\vert \hat \rho_0 \vert  \mathbf{q}_i' \rangle \, \langle \mathbf{q}_f' \vert \hat f \vert \mathbf{q}_f''\rangle. \nonumber \\ & \times & \sum_{\gamma} \vert A_{\gamma}(\mathbf{Q}_f \mathbf{Q}_i) \vert^2 \exp \left[ \frac{i}{\whbar} \left(\mathbf{p}^{(\gamma)}_f \mathbf{\delta q}_{f} - \mathbf{p}^{(\gamma)}_i \mathbf{\delta q}_{i} \right) \right] \nonumber  \\ & = & \int d\mathbf{Q}_i d\mathbf{Q}_f  \sum_\gamma \vert A_{\gamma}(\mathbf{Q}_f \mathbf{Q}_i) \vert^2 [\rho_0]_W ( \mathbf{Q}_i, \mathbf{p}^{(\gamma)}_i) \nonumber  \\ &  &   \hspace*{4.5cm} \times [f]_W ( \mathbf{Q}_f, \mathbf{p}^{(\gamma)}_f) \; ,
\end{eqnarray}
with $\hat \rho_0 \equiv \vert \Phi_0 \rangle \langle \Phi_0 \vert$ the initial density. Note that $ -{\partial^2 R_\gamma}/{\partial \mathbf{Q}_f \partial \mathbf{Q}_i} = {\partial \mathbf{P}_i}/{\partial \mathbf{Q}_f}_\gamma$. Therefore, the determinant in  Eq.~\eqref{eq:Agamma} is just the Jacobian of the transformation from the final ``position'' $\mathbf{Q}_f$ to the initial ``momentum''   $\mathbf{P}_i$.  Thus, $ \sum_\gamma \int d \mathbf{Q}_f \vert A_{\gamma} \vert^2 \mapsto (2\pi \whbar)^L \int d \mathbf{P}_i$ leading to Eq.~\eqrefEqTWA.


\begin{thebibliography}{41}
\bibitem [{rem()}]{rem_av}%
  \BibitemOpen
  \href@noop {} {}\bibinfo {note} {The time average is done by a convolution
  with a Gaussian of suitable width $\delta \tau$}\BibitemShut {NoStop}%
\end{thebibliography}
\end{document}